\documentclass{emulateapj}
\usepackage{apjfonts}

\begin{document}

\title{The detection of crystalline silicates in ultra-luminous infrared galaxies}

\author{
H.W.W. Spoon\altaffilmark{1,2},
A.G.G.M. Tielens\altaffilmark{3},
L. Armus\altaffilmark{4},
G.C. Sloan\altaffilmark{1},
B. Sargent\altaffilmark{5},
J. Cami\altaffilmark{6},
V. Charmandaris\altaffilmark{1,7,8},
J.R. Houck\altaffilmark{1},
B.T. Soifer\altaffilmark{4}
\email{spoon@astro.cornell.edu}
}

\altaffiltext{1}{Cornell University, Astronomy Department, Ithaca, NY 14853}
\altaffiltext{2}{Spitzer fellow}
\altaffiltext{3}{SRON National Institute for Space Research and
 Kapteyn Institute, P.O. Box 800, 9700 AV Groningen, The Netherlands}
\altaffiltext{4}{Caltech, Spitzer Science Center, MS 220-6, Pasadena, CA 91125}
\altaffiltext{5}{University of Rochester, Department of Physics and
 Astronomy, Rochester, NY 14627}
\altaffiltext{6}{NASA-Ames Research Center, MS 245-6, Moffett Field, CA 94035}
\altaffiltext{7}{Department of Physics, University of Crete,  
GR-71003, Heraklion, Greece}
\altaffiltext{8}{Chercheur Associ\'e, Observatoire de Paris, F-75014,  
Paris, France}

\slugcomment{Accepted for publication in the Astrophysical Journal, 2005 September 28}

\begin{abstract}
Silicates are an important component of interstellar dust and the
structure of these grains -- amorphous versus crystalline -- is sensitive
to the local physical conditions. We have studied the infrared spectra of
a sample of ultra-luminous infrared galaxies. Here, we report the
discovery of weak, narrow absorption features at 11, 16, 19, 23, and
28\,$\mu$m, characteristic of crystalline silicates, superimposed on the 
broad absorption bands at 10 and 18\,$\mu$m due to amorphous silicates in
a subset of this sample. These features betray the presence of forsterite
(Mg$_2$SiO$_4$), the magnesium-rich end member of the olivines. Previously,
crystalline silicates have only been observed in circumstellar environments. 
The derived
fraction of forsterite to amorphous silicates is typically 0.1 in these
ULIRGs. This is much larger than the upper limit for this ratio in the
interstellar medium of the Milky Way, 0.01. These results suggest that the
timescale for injection of crystalline silicates into the ISM is short in
a merger-driven starburst environment (e.g., as compared to the total time
to dissipate the gas), pointing towards massive stars as a prominent source
of crystalline silicates. Furthermore, amorphization due to cosmic rays,
which is thought to be of prime importance for the
local ISM, lags in vigorous starburst environments.
\end{abstract}



\section{Introduction}

Silicates are an important component of interstellar and circumstellar
dust. Many Galactic sources show broad emission or absorption features 
at 10 and 18 $\mu$m due to the Si--O stretching and O--Si--O bending 
vibrations of silicate bonds. Because of the width of these features, 
these silicates must have an amorphous structure.  These broad amorphous
silicate features have also been observed in a variety of
extragalactic environments, including Seyfert galaxies 
\citep[e.g.][]{laurent00,clavel00}, quasars \citep{hao05}, 
luminous and ultra-luminous infrared galaxies 
\citep{genzel98,rigopoulou99,tran01,spoon04,armus04} and a sample of 
mid-infrared detected, optically invisible, high-luminosity galaxies 
with redshifts of 1.7 $<$ z $<$ 2.8 \citep{houck05}.

In recent years, observational evidence for the presence
of {\it crystalline} silicates in various astrophysical environments
has also emerged. In particular, infrared spectra have revealed that
silicates in circumstellar environments often contain a significant 
crystalline fraction around both pre-main-sequence stars (T-Tauri
stars and Herbig AeBe stars) and post-main-sequence stars (Asymptotic
Giant Branch (AGB) stars, post-AGB stars, planetary nebulae, Luminous
Blue Variables (LBVs), Red Super Giants (RSGs) and post-RSGs)
\citep[e.g.][]{acke,malfait,molster02,sylvester,vanboekel,voors99,voors00}.
These crystalline silicates are invariably magnesium rich (e.g., 
pyroxene (MgSiO$_3$) and forsterite (Mg$_2$SiO$_4$).
Crystalline silicates are also known to be ubiquitous in the solar
system, including primitive objects such as comets 
\citep[e.g.][]{crovisier,hanner,wooden}. 
Because crystallization is 
inhibited by high-energy barriers, the origin and evolution of the 
crystalline silicate fraction in interstellar and circumstellar media 
has the potential to provide direct evidence of energetic
processing of grains. 

Silicates play an especially large role in shaping the mid-infrared 
spectral appearance of Ultra-Luminous InfraRed Galaxies (ULIRGs; 
L$_{\rm IR}\geq$10$^{12}$\,L$_{\odot}$). 
ULIRGs are thought to represent the final stage in the merging 
process of gas-rich spirals, where the interaction has driven
gas and dust towards the remnant nucleus, fueling a massive
starburst and a nascent active galactic nucleus (AGN).
Here, we present evidence for the presence of {\it crystalline}
silicates as part of the deep silicate absorption features observed 
towards a sample of twelve heavily obscured ULIRGs.


\section{Observations and data reduction}

We are obtaining mid-infrared spectroscopy for a sample of 110 ULIRGs,
as part of the Guaranteed Time Observation (GTO) ULIRG program of 
the Infrared Spectrograph
(IRS)\footnote{The IRS was a collaborative venture between Cornell
University and Ball Aerospace Corporation funded by NASA through the
Jet Propulsion Laboratory and the Ames Research Center}\citep{houck04}
on the Spitzer Space Telescope \citep{werner04}. Seventy seven of 
these spectra have been analyzed so far.

The twelve ULIRG spectra presented in this paper were selected based 
upon the strength of spectral structure indicative of crystalline 
silicate absorption bands at 16 and 23\,$\mu$m. Table \ref{tab1} lists
the basic properties of these targets, along with their observation
dates and on-source integration times.  

The observations were made with the Short-Low (SL) and Long-Low (LL)
modules of the {\em IRS}.  The spectra were extracted from the 
flatfielded images
provided by the Spitzer Science Center (pipeline version S11.0.2). The
images were background-subtracted by differencing the two SL apertures
and for LL, by differencing the two nod positions.  Spectra were then
extracted and calibrated using the {\em IRS} standard star HR\,6348 for
SL and the stars HR\,6348, HD\,166780, and HD\,173511 for LL
\citep{sloan05}. 
Small wavelength corrections were made to compensate for known offsets 
in the S11 processed data.
After extraction the orders were stitched to LL order 1, requiring 
order-to-order scaling adjustments of typically 5--10\%. 
The largest adjustment was made for Arp\,220 to match SL order 1 
and LL order 2, which required SL1 to be scaled up by 21\%.
In the final step, the 5--37\,$\mu$m spectra were scaled to match 
the observed {\em IRS} blue or red peak-up flux. For those sources 
lacking a (useful) peak-up flux, the spectra were scaled down by 
10\%, the average scaling factor of the other spectra.
In some spectra, most notably in IRAS\,20551--4250, residual fringing, 
which is due to subpixel pointing errors, appears in LL order 1 (between
20 and 30\,$\mu$m).

\begin{table} 
\caption{Properties of Sources\label{tab1}}
\begin{tabular}{lccccc}
\colrule
\colrule
Target     & AOR key & Date observed & Int. time & redshift & D$_L$ \\
           &         &               & min.      &          & Mpc\tablenotemark{a} \\
\colrule
00183--7111 & 7556352 & 14 Nov. 2003  & 15        & 0.327    & 1700       \\
00397--1312 & 4963584 & 04 Jan. 2004  & 14        & 0.262    & 1320       \\
01199--2307 & 4964864 & 18 Jul. 2004  & 14        & 0.156    & 737        \\
06301--7934 & 4970240 & 11 Aug. 2004  & 12        & 0.156    & 700        \\
06361--6217 & 4970496 & 11 Aug. 2004  & 14        & 0.160    & 760        \\
08572+3915  & 4972032 & 15 Apr. 2004  & 6         & 0.0584   & 258        \\
15250+3609  & 4983040 &  4 Mar. 2004  & 7         & 0.0554   & 244        \\
Arp220      & 4983808 & 29 Feb. 2004  & 5         & 0.0181   & 78         \\
17068+4027  & 4986112 & 16 Apr. 2004  & 14        & 0.179    & 858        \\
18443+7433  & 4987904 &  5 Mar. 2004  & 14        & 0.135    & 627        \\
20551--4250 & 4990208 & 14 May  2004  & 4         & 0.0427   & 186        \\
23129+2548  & 4991488 & 17 Dec. 2003  & 23        & 0.179    & 858        \\
\colrule
\end{tabular}
\tablenotetext{a}{assuming H$_0$=71\,km\,s$^{-1}$\,Mpc$^{-1}$, 
$\Omega_M$=0.27, $\Omega_{\Lambda}$=0.73, $\Omega_K$=0}
\end{table}

\begin{figure}
\begin{center}
\resizebox{\hsize}{!}{\includegraphics{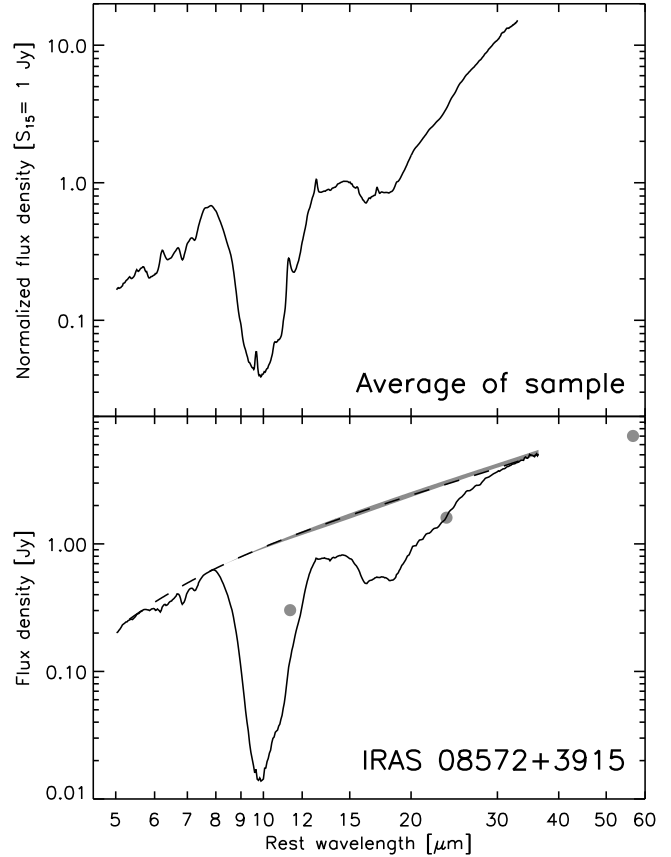}}
\caption{Upper panel: the average 5--32\,$\mu$m spectrum of the 12 ULIRGs 
in our sample after normalizing each spectrum individually at
15\,$\mu$m. Features and lines are identified in the text.
Lower panel: the 5--60\,$\mu$m spectrum of IRAS\,08572+3915. The IRS
low-resolution spectrum ({\it black}) is shown together with the
12, 25 and 60\,$\mu$m IRAS fluxes ({\it grey filled circles}). 
Three choices for the local continuum are indicated by the 
{\it black dashed line} and {\it grey shaded area}. Error bars
are ommited in both panels, because of the high S/N of the data
\label{fig1}}
\end{center}
\end{figure}


\section{Analysis}

The top panel of Figure\,\ref{fig1} shows the average {\em IRS} 
low-resolution spectrum for our sample, obtained after scaling the 
spectra to S$_{\nu}$\,=\,1\,Jy at 15\,$\mu$m. The average
spectrum is dominated by broad absorption features at 10 and
18\,$\mu$m, which we attribute to absorption by amorphous silicates.
The spectrum further shows weaker absorption features due to water ice 
(6.0\,$\mu$m) and hydrocarbons (6.90 and 7.25\,$\mu$m). PAH emission 
features can be seen at low contrast at 6.2, 7.7, and 11.2\,$\mu$m.
The features are especially weak in the spectrum of IRAS\,F00183--7111 
\citep{spoon04} and absent in the spectrum of IRAS\,08572+3915
(Figure\,\ref{fig1}, lower panel). Our spectra further show
a few emission lines, most notably the H$_2$\,S(3) line at
9.66\,$\mu$m, the [Ne{\sc ii}] line at 12.81\,$\mu$m and the 
H$_2$\,S(1) line at 17.0\,$\mu$m. These will be discussed elsewhere 
\citep{armus05,higdon05}. 

Upon close inspection, the spectra also reveal evidence for weak 
and narrow absorption features near 16, 19 and 23 $\mu$m, independent 
of the redshift of the source. Emission and absorption features at 
these wavelengths are also know in Galactic sources with strong 
silicate features and have been attributed to the presence of 
crystalline silicates.

\begin{figure}
\begin{center}
\resizebox{\hsize}{!}{\includegraphics{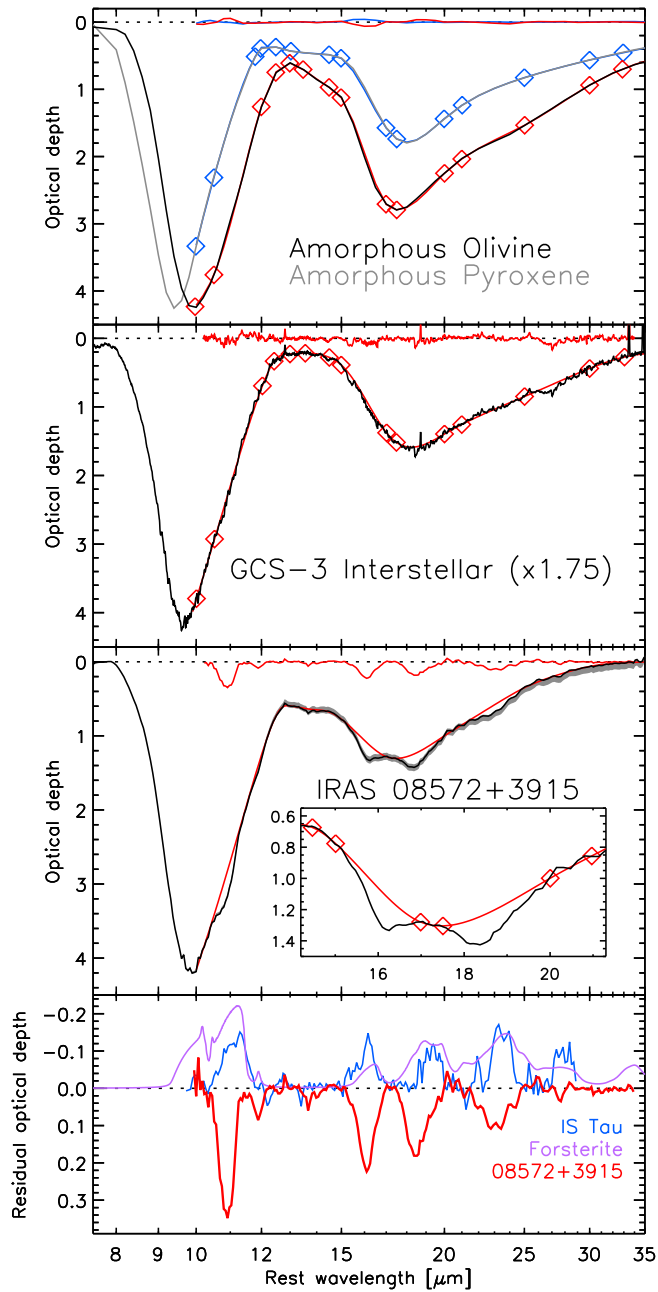}}
\caption{Illustration of our method for fitting the amorphous 
component of silicate absorption features.
Upper panel: laboratory opacity spectra of amorphous olivine
({\it black}) and amorphous pyroxene ({\it grey}), presented 
as optical depth profiles. Overlayed are spline fits 
({\it red, blue}) to pre-selected wavelength pivots 
({\it red, blue diamonds}). 
Fit residuals, indicating imperfections in the fitting method,
are shown as deviations from zero optical depth ({\it red, blue}).
Second panel: spline fit ({\it red}) to the GCS\,3 optical 
depth profile ({\it black}; scaled-up by a factor 1.75). Fit 
residuals are shown as deviations from zero optical depth 
({\it red}).
Third panel: the optical depth spectrum of IRAS\,08572+3915 
({\it black}). The {\it grey shading} indicates the effect of a 
different choice of local continuum on the resulting optical 
depth profile; see Figure\,\ref{fig1}. The {\it red} curve is 
a spline fit to the amorphous silicate profile, ignoring the 
narrow substructure near 11, 16, 19 and 23\,$\mu$m. The residual
spectral structure, shown in {\it red} at the top of this panel, 
is too strong to be attributed to imperfections in the fitting
method. Instead we attribute the features at 11, 16, 19 and 
23\,$\mu$m to crystalline silicates.
Third panel inset: a close-up of the 14--21\,$\mu$m range, 
demonstrating the prescribed placement of spline pivots at 
14.5, 15.0, 17.0, 17.5, 20.0 and 21.0\,$\mu$m.
Lower panel: the residual optical depth spectrum of IRAS\,08572+3915 
({\it red} is compared to the residual crystalline silicate emissivity 
of the disk around the young star IS\,Tau ({\it blue}) and the opacity 
profile of forsterite ({\it purple}). Both have been scaled by
arbitrary factors
\label{fig2}}
\end{center}
\end{figure}

\begin{figure*}
\begin{center}
\resizebox{\hsize}{!}{\includegraphics{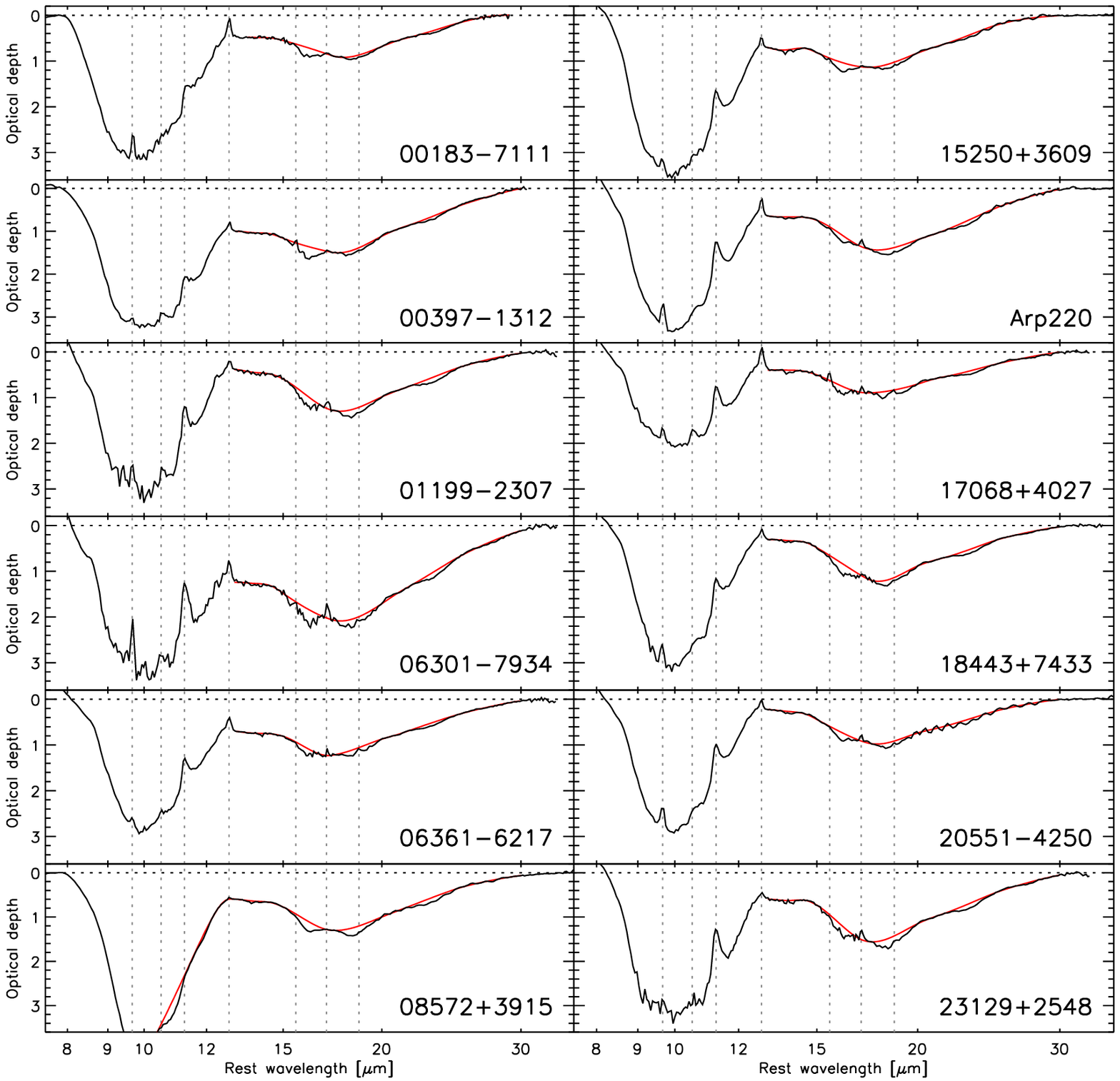}}
\caption{8--30\,$\mu$m optical depth spectra for the 12 ULIRGs in our 
sample. The {\it red} curve is a spline fit to the amorphous silicate 
profile, ignoring the narrow substructure at 11, 16, 19, 23 and 
28\,$\mu$m. 
Vertical dotted lines indicate the positions of 9.66\,$\mu$m 
H$_2$ S(3), 10.5\,$\mu$m [S{\sc iv}], 11.3\,$\mu$m PAH, 
12.8\,$\mu$m [Ne{\sc ii}], 15.5\,$\mu$m [Ne{\sc iii}], 
17.0\,$\mu$m H$_2$ S(1) and 18.7\,$\mu$m [S{\sc iii}].
\label{fig3}}
\end{center}
\end{figure*}

In order to investigate the presence of a crystalline component to
the silicate absorption features in our sample, we infer the silicate 
optical depth spectrum by dividing out a local continuum from our
{\em IRS} spectra. We define the local continuum as a three point 
spline interpolation of continuum points at 5.6\,$\mu$m (0.1\,$\mu$m
shortward of the wavelength range affected by the 6.0\,$\mu$m water 
ice absorption feature), 7.1\,$\mu$m (in between the two hydrocarbon
absorption bands at 6.90 and 7.25\,$\mu$m), and the red cut-off of 
{\rm IRS} LL order 1 (ranging from 29 to 36\,$\mu$m, depending on 
the redshift of the source). 
For sources without discernable 7.7\,$\mu$m PAH emission
(e.g. IRAS\,08572+3915 and IRAS\,F00183--7111), we replace the 
7.1\,$\mu$m pivot by the continuum at 7.9--8.0\,$\mu$m.
The resulting spline interpolated local continuum 
is illustrated for IRAS\,08572+3915 in the lower panel of 
Figure\,\ref{fig1}.
The effect of small changes in the adopted local continuum on
the resulting silicate profile is illustrated by the shaded areas 
around the spline-interpolated continuum in the lower panel of 
Figure\,\ref{fig1} and on the resulting silicate profile in the 
third panel of Figure\,\ref{fig2}. 
The resulting optical depth profiles for the twelve sources in 
our sample have apparent 10\,$\mu$m optical depths ranging from
2.1 for IRAS\,17068+4027 to 4.2 for IRAS\,08572+3915, with a mean
value of 2.5 (see Figure\,\ref{fig3} and Table\,\ref{tab2}).

In order to derive the characteristics of these narrow features we have 
to account for the amorphous silicate features which dominate the 
absorption structure in this wavelength range.  We have compared the 
general profile of the 10 and 18\,$\mu$m amorphous silicate featues in 
these ULIRGS with those in a sample of Galactic background sources 
\citep{chiar05}. 
However, because of the uncertainties in the placement of
the continuum for these Galactic sources, the derived interstellar
10 and $18\mu$m amorphous silicate features show differences
in their relative strengths, and in the level of absorption between 
them.
Since fitting the ULIRG data with a composite Galactic spectrum
would therefore introduce spurious artifacts, we have elected to
represent the broad amorphous silicate features in the ULIRG
and Galactic spectra with spline fits.  Deviations from these fits
in the ULIRG spectra, significantly larger than those seen among
the Galactic sources, would then argue strongly for a crystalline
component. We have tested this procedure on the high signal-to-noise
spectrum of the Galactic Center Source GCS\,3, and on laboratory
absorption spectra of amorphous silicates. The residuals from
the spline fits to these spectra are very small ($<$6\%), validating
our analysis to that level.

The method is demonstrated in the top panel of Figure\,\ref{fig2}
for laboratory spectra of amorphous olivines and pyroxenes 
\citep{fabian01}. The spline fit traces the amorphous profile to 
better than 2\% of the local optical depth, except for the 16\,$\mu$m 
range, where the spline deviates by up to 6\%. The resulting 
small artifacts are shown at the top of the panel.
We also tested our method on the optical depth profile as seen 
towards the Galactic background source GCS\,3 \citep{chiar05},
depicted in the second panel of Figure\,\ref{fig2}. The silicates 
along this line of sight are thought to be $>$99\% amorphous in 
composition \citep{kemper04,kemper05}. In contrast to the laboratory 
profiles, the fit residuals for the GCS\,3 spectrum are dominated 
by spectral noise rather than fitting artifacts.
The third panel of Figure\,\ref{fig2} shows the fit to the silicate 
profile of the ULIRG IRAS\,08572+3915. The red curve smoothly fits 
the maximum depth of the broad 18\,$\mu$m feature, while ignoring 
the strong, narrow substructure around 11, 16, 19 and 23\,$\mu$m. 
After subtracting the spline-interpolated amorphous component from 
the optical depth spectrum, the features at 11, 16, 19 and 23\,$\mu$m
show up in the fit residual spectrum, clearly above the levels
expected for artifacts introduced by the fitting procedure.
In the lower panel of Figure\,\ref{fig2} we compare these residuals 
to known spectra of circumstellar and laboratory crystalline 
silicates.

\begin{figure*}
\begin{center}
\resizebox{13cm}{!}{\includegraphics{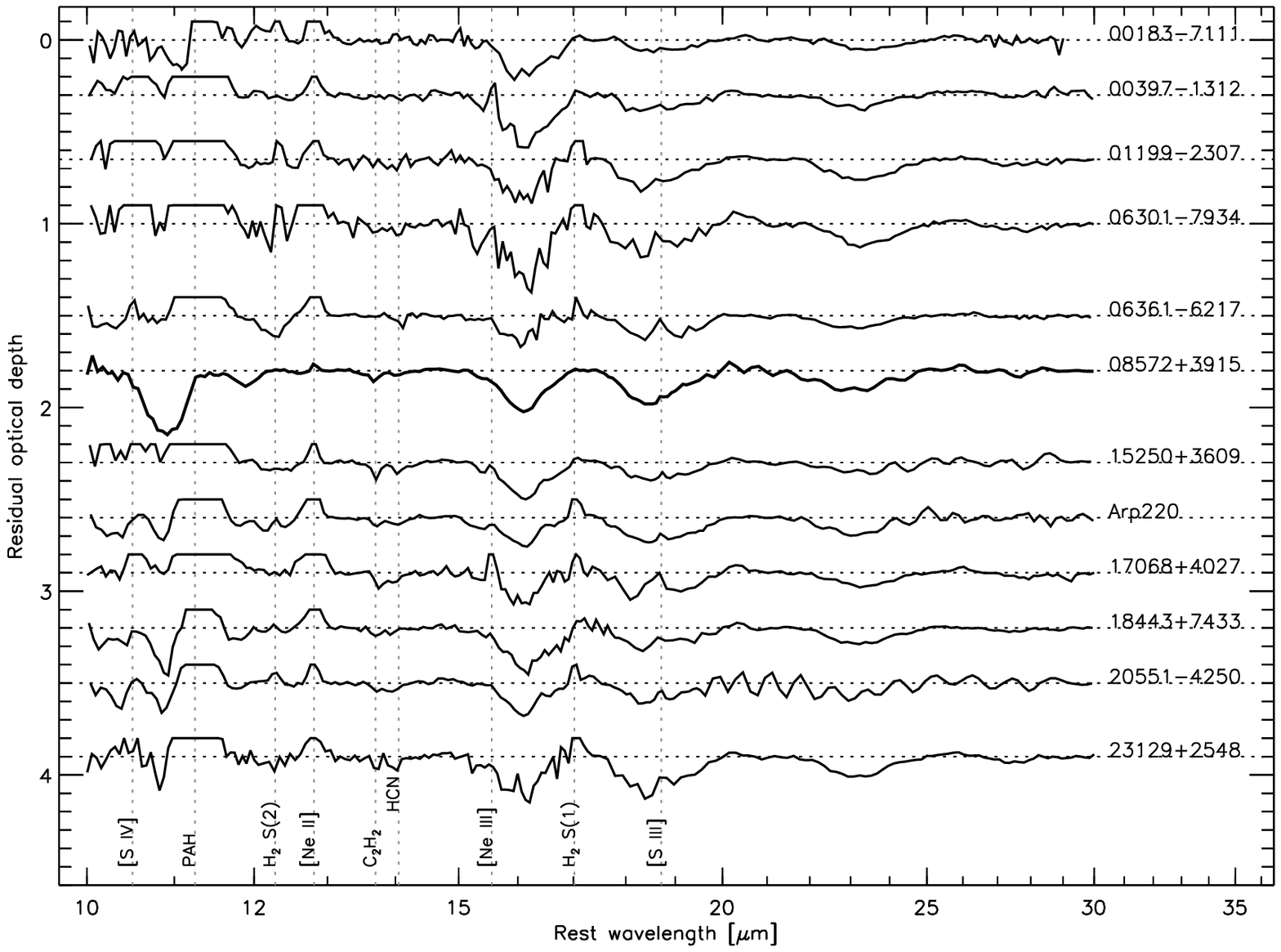}}
\caption{10--30\,$\mu$m residual optical depth spectra for the 12
ULIRGs in our sample after subtraction of the amorphous silicate 
component from the optical depth spectra. For plotting purposes,
the spectra have been offset and truncated at residual optical 
depths of -0.1. Crystalline silicate features can be identified 
at 11, 16, 19, 23 and 28\,$\mu$m
\label{fig4}}
\end{center}
\end{figure*}

The adopted spline fits to the amorphous silicate profiles for 
the twelve ULIRGs in our sample are overplotted in red in 
Figure\,\ref{fig3}, while the residual optical depth spectra 
are presented in Figure\,\ref{fig4}. The individual spectra 
in the latter figure are truncated at -0.1 optical 
depth in order not to be dominated by spurious optical depth 
structure introduced by emission features of 11.3\,$\mu$m PAH 
and 12.8\,$\mu$m [Ne {\sc ii}]. 
These common features aside, the twelve spectra clearly show 
absorption structure at 16, 19 and 23\,$\mu$m that cannot be 
attributed to observational artifacts (i.e. fringing, flux 
calibration) or imperfections in the spline fit to the 
amorphous component. 
A few sources (e.g. IRAS\,06301--7934, IRAS\,23129+2548) 
further show a weak feature at 27--28\,$\mu$m and only one source, 
IRAS\,08572+3915, a strong feature at 11\,$\mu$m (see also 
Figure\,\ref{fig2}). The latter feature may also be present
in the other spectra, but its detection is complicated by the
presence of emission from 10.51\,$\mu$m [S {\sc iv}]
and 11.3\,$\mu$m PAH in the same wavelength range
(see Figure\,\ref{fig4}).
Further note the presence of absorption bands of gas phase
C$_2$H$_2$ and HCN at 13.7 and 14.05\,$\mu$m in the spectra 
of several of our ULIRGs, most notably IRAS\,15250+3609 and 
IRAS\,23129+2548. These bands are otherwise seen in dense, 
warm molecular clouds surrounding massive Galactic protostars 
\citep{lahuis00}.

\begin{figure}
\begin{center}
\resizebox{\hsize}{!}{\includegraphics{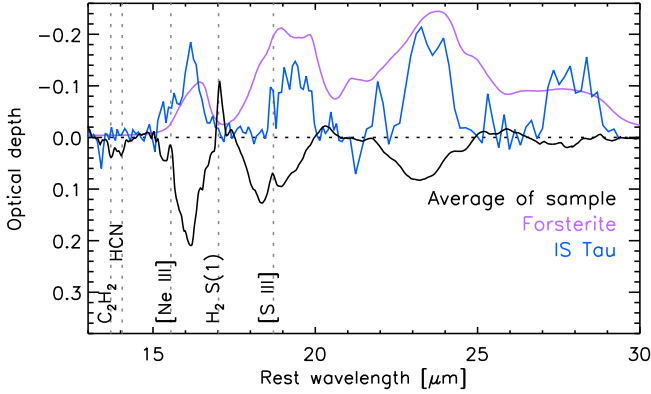}}
\caption{The average 13--30\,$\mu$m residual optical depth spectrum
for our sample ({\it black}) compared to the residual crystalline 
silicate emissivity of the disk around the young star IS\,Tau 
({\it blue}) and the opacity profile of forsterite ({\it purple}).
Both have been scaled by arbitrary factors
\label{fig5}}
\end{center}
\end{figure}

In Figure\,\ref{fig5} we show the average residual optical depth
spectrum for our twelve sources and compare it to the observed
crystalline silicate opacity profile of the young star IS Tau 
\citep{sargent05} and the calculated opacity profile of forsterite
\citep{fabian01}. 
The relative strengths of the absorption bands in the ULIRG spectra
are different from those of the comparison profiles. In the ULIRG
spectra, the features get progressively weaker towards longer 
wavelength, whereas no such trend exists for the comparison profiles
(see also Figures \ref{fig2} and \ref{fig5}). 
A possible explanation for this trend could be dilution of the
crystalline absorption spectrum by cold dust emission from a less
obscured cooler component, either the other nucleus or the 
circumnuclear environment.
Nevertheless, the similarities between the ULIRG profile and 
the comparison profiles in peak position and width are striking. 
We therefore conclude that the features observed at 11, 16, 19, 23 
and 28\,$\mu$m in the ULIRG spectra are indeed caused by the 
presence of crystalline silicates in their nuclear medium. 

The peak position of the 16\,$\mu$m band, in particular, is sensitive 
to the Mg/Fe ratio of olivines \citep{koike03}. The peak position of 
this band observed in the ULIRG spectra (16.1\,$\mu$m) falls close to 
the laboratory measured position of forsterite, the Mg-rich end member of
the olivines \citep[Mg$_2$SiO$_4$; 16.3\,$\mu$m;][]{koike03} and much to
the blue of the peak position in fayalite, the iron-rich end member of
the olivines \citep[Fe$_2$SiO$_4$; 17.7\,$\mu$m;][]{koike03}. 
Hence, in line with other observations of crystalline silicates in
space, extragalactic olivines appear to be extremely Mg-rich and Fe-poor.

\begin{table} 
\caption{Amorphous and crystalline silicates \label{tab2}}
\begin{tabular}{lcccc}
\colrule
\colrule
Target      & $\tau_{am}$(peak)\tablenotemark{a} & $\tau_{cr}$(peak)\tablenotemark{b} 
            &  $\tau_{cr}$(peak)\,/$\tau_{am}$(peak) & N$_{cr}$\,/N$_{am}$\tablenotemark{c} \\
\colrule
00183--7111 & 3.0   &    0.22   &  0.073   & 0.11 \\
00397--1312 & 3.2   &    0.28   &  0.088   & 0.13 \\
01199--2307 & 3.6   &    0.21   &  0.060   & 0.09 \\
06301--7934 & 3.4   &    0.35   &  0.10    & 0.15 \\
06361--6217 & 2.9   &    0.15   &  0.051   & 0.08 \\
08572+3915  & 4.2   &    0.22   &  0.053   & 0.08 \\
15250+3609  & 3.6   &    0.20   &  0.056   & 0.08 \\
Arp220      & 3.3   &    0.16   &  0.048   & 0.07 \\
17068+4027  & 2.1   &    0.17   &  0.082   & 0.12 \\
18443+7433  & 3.2   &    0.25   &  0.078   & 0.12 \\
20551--4250 & 2.9   &    0.18   &  0.062   & 0.09 \\
23129+2548  & 3.2   &    0.25   &  0.078   & 0.12 \\
\colrule
\end{tabular}
\tablenotetext{a}{apparent peak optical depth of the 10\,$\mu$m feature}
\tablenotetext{b}{apparent peak residual optical depth of the
  16\,$\mu$m feature}
\tablenotetext{c}{assuming N$_{cr}$/N$_{am}$ = 1.5 $\times$
  $\tau_{cr}$(peak)\,/$\tau_{am}$(peak) ; see eq. (1)}
\end{table}

We infer the fraction of crystalline silicates in our sample 
from the peak optical depths of the 10\,$\mu$m amorphous and 
the 16\,$\mu$m crystalline silicate absorption bands using
\begin{equation}
\frac{N_{cr}}{N_{am}} = \frac{\tau_{cr}({\rm peak})}{\tau_{am}({\rm peak})} \times 
\frac{\kappa_{am}({\rm peak})}{\kappa_{cr}({\rm peak})}
\end{equation}
where $N_{am}$ and $N_{cr}$ are the mass column densities of the
amorphous and crystalline silicates.
Adopting $\kappa_{am}$(peak) = 2.4$\times$10$^3$ cm$^2$/g as the peak mass 
absorption coefficient for amorphous silicates \citep{dorschner95}
and $\kappa_{cr}$(peak) = 1.6$\times$10$^3$ cm$^2$/g as the peak mass 
absorption coefficient for forsterite \citep{fabian01}, we find 
crystalline-to-amorphous silicate mass column density ratios ranging 
from 0.07 to 0.15, with a median value of 0.11 (see Table\,\ref{tab2}). 
It should be understood that these numbers are upper limits,
since foreground emission and radiative transfer effects within 
the optically thick 10\,$\mu$m silicate band may cause the apparent 
optical depth of the 10\,$\mu$m silicate feature to be a lower 
limit to the true optical depth.
Our findings should be contrasted to the upper limit of crystalline 
silicates in the general ISM of the Milky Way of $<$1\% 
\citep{kemper04,kemper05} and the crystalline-to-amorphous ratios of
up to 0.75 observed in circumstellar environments \citep{molster02}.

Given the broad range in infered crystalline-to-amorphous silicate 
ratios in our sample, we have investigated the existence of 
correlations with the apparent 10\,$\mu$m optical depth, the 
IRAS R(60,100) color, the 5.5\,$\mu$m--to--23\,$\mu$m rest frame 
spectral slope and the 6.2\,$\mu$m PAH equivalent width, but found no 
significant trends. The fraction of crystallinity within this sample
can therefore, at present, not be linked to the infrared spectral 
appearance.


\section{Discussion}

We have detected crystalline substructure in the silicate absorption 
features towards twelve strongly obscured ULIRG nuclei. This is the first
detection of crystalline silicates in any source outside the Local Group. 
Space-based and ground-based observations have revealed the presence
of infrared emission features of crystalline silicates around young
stellar objects and evolved stars  
\citep{acke,malfait,molster02,sylvester,vanboekel,voors99,voors00}.
Comets also often show evidence for crystalline silicates in their spectra 
\citep{crovisier,hanner,wooden}. In contrast, while stellar sources
of silicate dust inject at least 5\% of their silicates
in crystalline form, the Galactic {\it interstellar} silicate feature is
exceedingly smooth, and there is no evidence for a crystalline absorption 
component in the Galactic ISM. This translates into an upper limit on 
the crystalline fraction in the Galactic interstellar medium of 
1\% \citep{kemper04,kemper05}. This difference in crystallinity 
between silicates injected and silicates in the ISM implies a rapid 
transformation of crystalline silicates into amorphous silicates in
the interstellar medium \citep[$\la 10^8$\,yr;][]{kemper04,kemper05}. 
This transformation has been attributed to energetic processing of 
the dust by heavy cosmic ray ions \citep{bringa} or to ion bombardment 
in high velocity (v\,$>$\,1000\,km/s) shocks \citep{jaeger,rotundi,demyk}. 
In particular, based upon laboratory experiments and estimated cosmic 
ray fluxes in our galaxy, the timescale for amorphization is estimated 
to be only 70 million years \citep{bringa}, considerably shorter 
than the timescale at which (crystalline) silicates are injected into 
the Galactic ISM \citep[4 billion years;][]{bringa}

These same processes --- injection of crystalline silicates from stars 
and cosmic ray and shock amorphization --- will play a role in the
interstellar media of these ULIRGs as well. Possible explanations for
the much higher crystalline silicate fraction in these ULIRGs as
compared to the Milky Way include a higher fraction of crystalline
silicates injected and/or a delay in the interstellar amorphization
rate. 

In particular, ULIRGs are characterized by a high rate of star
formation driven by merging events. In contrast to the local ISM, the
enrichment of the dusty interstellar medium will be dominated by the
rapidly evolving massive stars and the contribution by the more
numerous low-mass stars will lag on the timescale associated with the
ULIRG-merger event \citep[$10^8-10^9$ yr;][]{murphy96,murphy01}. 
Evidence for the presence of a large population of evolved massive 
stars in ULIRG nuclei exists for the nearest ULIRG, Arp\,220 
\citep{armus95}.
Spectroscopic studies of sources in our galaxy have revealed that 
massive stars are prominent sources of crystalline silicates during 
the supergiant phase. In particular, the more extreme examples of 
this class, such as AFGL\,4106, NML\,Cyg and IRC\,+10420, have 
crystalline silicate fractions of the order of 0.15 \citep{molster99}. 
Likewise, crystalline silicates have been observed in the ejecta 
of some LBVs, such as R\,71 and AG\,Car \citep{voors99,voors00}. 
This episodic mass-loss phase dominates the dust injection by the 
most massive stars (M$>$50M$_{\odot}$) in the Galaxy. 
Finally, it is currently unknown how much crystalline silicate dust 
is injected by supernovae. Indeed it is unknown how much dust is 
injected by SNe. However, it is
conceivable that the SNe resulting from this starburst also inject a
large fraction of their freshly condensed dust in the form of
crystalline silicates. Hence, a starburst may rapidly increase the
crystalline silicate fraction in the nuclear region.

Of course, the high supernova rate in a starburst will drive 
strong shock waves into the local environment, which will sputter 
the dust, returning the atoms from the solid to the gas phase  
\citep{jones94,jones96}. However, likely, the destruction rate of 
crystalline and amorphous silicates are similar and hence, to
first order, this will not affect the crystalline-to-amorphous
fraction.  More importantly, supernovae are thought to be the dominant
source of cosmic rays and this may increase the rate at which
crystalline silicates are transformed into amorphous silicates. 
However, cosmic rays freshly accelerated in the supernova remnants
may leak away to the rest of the galaxy on a timescale of 
$\sim$10$^6$ yr \citep{webber93}, limiting the rate of this 
solid-state conversion in the starburst environment.  
Thus, the crystalline-to-amorphous conversion time-scale
may actually be quite similar to the value calculated for the local
Milky Way \citep[70 million years;][]{bringa}. We expect that over
time, as the merging event ages and the starburst intensity 
decreases, the fraction of crystalline silicates will decrease 
on a similar time-scale. Thus, we attribute the high fraction 
of crystalline silicates in these ULIRGs as compared to others 
to the relative `youth' of these systems; e.g., the amorphitization
process may lag the merger-triggered, star-formation-driven dust 
injection process.

All 17 ULIRGs with $\tau$(10\,$\mu$m)\,$>$\,2.9 in our ULIRG sample 
of 77 galaxies show crystalline silicate structure in their IRS spectra.
At lower silicate optical depth, the number of clear detections falls
sharply. Above $\tau$(10\,$\mu$m)=2, 18 out of 23 ULIRGs
definitely show the 16\,$\mu$m crystalline feature. Above 
$\tau$(10\,$\mu$m)=1, this is 20 out of 46 and for the 
entire ULIRG sample only 21 out of 77. 
We note that a complete analysis of this fraction is hampered by 
possible radiative transfer effects in the silicate features
(as evidenced by the strong variation in the ratio of
10--to--18\,$\mu$m peak optical depths), 
the presence of 16--18\,$\mu$m UIR emission in some sources with 
non-negligible PAH emission \citep{brandl05} and by the challenge 
to detect weak 16\,$\mu$m absorption features at low contrast and 
low signal-to-noise. Nevertheless, it is clear that crystalline 
silicates are a common component of the nuclear interstellar medium
during the strongly obscured evolutionary phase of the ULIRG phenomena.

Crystalline silicates can also be a sign of high-temperature grain
processing.  In a ULIRG, both a central AGN and hot, young stars
are potential sources of high-energy photons capable of heating
the dust in the ISM.  X-ray and far UV photons from a central AGN
will produce a hot, inner region.  However, an AGN as the source
of grain processing in our ULIRG sample is problematic for two
reasons. First, interferometric  10\,$\mu$m spectra of the
Seyfert-2 nucleus of NGC\,1068 do not show any indication for 
crystalline silicates in its inner warm ~2\,pc region \citep{jaffe04}.
Second, the crystalline silicates are seen in absorption in our 
spectra, implying that they are located in the cooler, outer 
regions. This would require a large scale mechanism to transport
the inner toroid material outwards and distribute it over the 
surrounding medium. The extended nature of a starburst does not 
require such a mechanism and therefore seems a more likely source 
of processed material than a central source.

Crystalline silicates are also a characteristic of the circumstellar 
planetary disks surrounding young stellar objects such as 
Herbig AeBe stars and T-Tauri stars. Again, in order to be seen in 
absorption, this material will have to be transported outwards into 
the surrounding cold medium, presumably through jets and winds 
\citep{tielens03}. There is, however, no indication in Galactic 
sources for such large scale transport of crystalline silicates. 
On the contrary, for Herbig AeBe stars the emission features 
of crystalline silicates are strongly concentrated towards the 
inner 2\,AU of the disks \citep[e.g.,][]{vanboekel}. Hence, we deem 
high-temperature crystallization of existing amorphous silicate 
grains unlikely as the source of the high crystalline
silicate fraction in ULIRGs.

\section{Conclusions}
In this work, we report the discovery of crystalline substructure 
at 11, 16, 19, 23 and 28\,$\mu$m in the amorphous silicate features 
towards twelve deeply obscured ULIRG nuclei. These features indicate
the presence of the mineral forsterite (Mg$_2$SiO$_4$). 
Previously, crystalline silicates have only been observed in 
circumstellar environments.

We infer the fraction of crystalline silicates in our sample from
the peak optical depths of the 10\,$\mu$m amorphous and the 
16\,$\mu$m crystalline bands and find a crystalline-to-amorphous
ratio ranging from 0.07 to 0.15, with a median value of 0.11. These
numbers are likely upper limits, since foreground emission and
radiative transfer effects will cause the apparent 10\,$\mu$m
silicate optical depth to be a lower limit to the true optical
depth.

The crystalline-to-amorphous ratio in our twelve deeply obscured
ULIRGs is 7--15 times larger than the upper limit for this ratio 
in the interstellar medium of the Milky Way. This suggest that the 
timescale for injection of crystalline silicates into the ISM is 
short in a merger-driven starburst environment (eg., as compared 
to the total time to dissipate the gas), pointing towards evolved
massive stars (red supergiants, LBVs and type {\sc ii} supernovae) 
as prominent sources of crystalline silicates. 
Furthermore, the timescale for amorphitization of crystalline 
silicates, which is known to be fairly rapid in the ISM of the
Milky Way ($\sim$\,10$^8$ yr), is at most of similar order in 
starburst environments.
We expect that over time, as the merging event ages and the 
starburst decreases in intensity, the fraction of crystalline 
silicates will decrease rapidly. Thus, we attribute the high 
fraction of crystalline silicates in these ULIRGs, as compared 
to others, to the relative `youth' of these systems.

Finally, other galaxy types may also exhibit crystalline silicate
features far above the upper limits set for the crystalline silicate 
fraction in the ISM of our galaxy \citep[$<$1\%;][]{kemper04,kemper05}.
The possible detection of a 23\,$\mu$m crystalline emission feature 
in the quasar PG\,1351+640, reported by \cite{hao05}, may be a signpost 
of more detections to come.


\acknowledgments

The authors wish to thank Elise Furlan, Bill Forrest, 
Patrick Morris, Els Peeters for discussions, and Frank Molster 
and Sacha Hony for sharing their ISO--SWS data. 
Support for this work was provided by 
NASA through Contract Number 1257184 issued by the Jet Propulsion 
Laboratory, California Institute of Technology under NASA contract 
1407.  HWWS was supported under this contract through the 
Spitzer Space Telescope Fellowship Program.

\end{document}